\documentclass[11pt,draftcls,onecolumn]{IEEEtran}

\usepackage{graphicx}
\usepackage{subcaption}
\usepackage{textcomp}

\begin{document}

\title{Augmenting Bottleneck Features of Deep Neural Network Employing Motor State for Speech Recognition at Humanoid Robots}

\author{Moa~Lee
        and~Joon-Hyuk~Chang,~\IEEEmembership{Senior~Member,~IEEE}
}
\maketitle

\begin{abstract}
As for the humanoid robots, the internal noise, which is generated by motors, fans and mechanical components when the robot is moving or shaking its body, severely degrades the performance of the speech recognition accuracy. In this paper, a novel speech recognition system robust to ego-noise for humanoid robots is proposed, in which on/off state of the motor is employed as auxiliary information for finding the relevant input features. For this, we consider the bottleneck features, which have been successfully applied to deep neural network (DNN) based automatic speech recognition (ASR) system. When learning the bottleneck features to catch, we first exploit the motor on/off state data as supplementary information in addition to the acoustic features as the input of the first deep neural network (DNN) for preliminary acoustic modeling. Then, the second DNN for primary acoustic modeling employs both the bottleneck features tossed from the first DNN and the acoustics features. When the proposed method is evaluated in terms of phoneme error rate (PER) on TIMIT database, the experimental results show that achieve obvious improvement (11\% relative) is achieved by our algorithm over the conventional systems.
\end{abstract}


%
\IEEEpeerreviewmaketitle

\section{Introduction}


The automatic speech recognition (ASR) technology, the most natural and intuitive means of communication for human-robot interaction, becomes more essential because humanoid robots perform actions or responds according to human commands. Many humanoid robots or similar robots, in reality, including the Softbank\textquotesingle s robot Pepper \cite{Pepper}, the MIT\textquotesingle s home robot JIBO \cite{JIBO}, and Intel\textquotesingle s Jimmy \cite{Jimmy} have been developed based on the ASR technology. Recently, many researches on this ASR technology, as an indispensable part for the humanoid robots, have been actively carried out, but still remains a challenging problem The humanoid robots, especially, generate strong internal noise, which results in a significant factor deteriorating the recognition performance. Indeed, because of the close distance between the microphone and the motor or joint than the human voice source, the internal noise incurred from motors, fans, and mechanical components noise is loudly recorded into the microphone installed on the robot especially while the robot is actively moving This self-created noise in humanoid robot is referred to as ego-noise \cite{Ince}, which has been not been fully treated to be addressed while the robust speech recognition in the presence of background noise or external interference has been extensively studied thus far \cite{ego-noise1}-\cite{MOA}.


As for ego-noise suppression, spectral subtraction \cite{SS} is one of the common methods. For instance, Ito \textit{et al.} \cite{ego-noise1} developed an framewise prediction approach based on a neural network (NN), where the noise spectra are predicted by using angular velocities of the joints of the robot. Then, the estimated noise spectra are subtracted from the target signal spectra. One of the problem in this approach is that the ASR performance quite becomes poor when the noise power is not well-estimated especially when the ego-noise is non-stationary. Several researchers have tackled this problem by predicting and subtracting ego-noise using templates. Nishimura \textit{et al.} \cite{ego-noise5} proposed a method to predict the ego-noise using motion like gesture and walking pattern template obtained from a pre-recorded motor noise corresponding to the motion pattern. With the labeled motion command, the appropriate ego-noise template matched to latest motion is selected from the template database and used for subtraction.


Ince \textit{et al.} \cite{Ince} extended the small set of noise template database to larger ego-noise space in which the template database was enhanced by incorporating more information related to the joints such as angular positions, velocities and accelerations. Schmidt \textit{et al.} \cite{ego-noise4} employed the motor data to predict the intrinsic harmonic structure of ego-noise and incorporate the ego-noise harmonics into a multichannel dictionary-based ego-noise reduction approach. These studies on ego-noise reduction show that, unlike the conventional background noise suppression method, the instantaneous motor data of the humanoid robot can be used as a secondary information source for dealing with the ego-noise problem. Recently, we originally devised an idea in \cite{MOA} to use the motor on/off state as the auxiliary information when designing the acoustic model of the deep neural network (DNN)-based speech recognition system for humanoid robots. However, the auxiliary information is simply designed as a one-hot vector, so the performance gain is limited.


In this paper, we propose a new approach based on the bottleneck features to further improve the speech recognition performance when using the on/off motor state as auxiliary information A first DNN is carefully designed to create motor state dependent bottleneck features for which the first DNN input is determined by concatenating the motor on/off state data in addition to the acoustic features contaminated by the ego-noise. Then, preliminary training is accomplished to yield the bottleneck features, which are fed into the second DNN input, designed for primary acoustic modeling under ego-noise environments. Finally, the second DNN is trained with the input including both the bottled features and acoustic features to fully represent the complex relationship between the audio signal and phoneme. In order to verify the performance of our new approach against the existing methods, experiments are extensively conducted on TIMIT corpus. The experimental results showed better performance in terms of phoneme error reduction (PER) reduction than the baseline models including \cite{MOA} and the method that used only acoustic features.

The rest of this paper is organized as follows. The related works and the proposed methods are described in Section II and Section III, respectively. Section IV presents the experimental setting and shows results. Then, Section V concludes the work.



\section{Bottleneck features}

\begin{figure*}[tbh]
    \centering
    \includegraphics[height=2.8in]{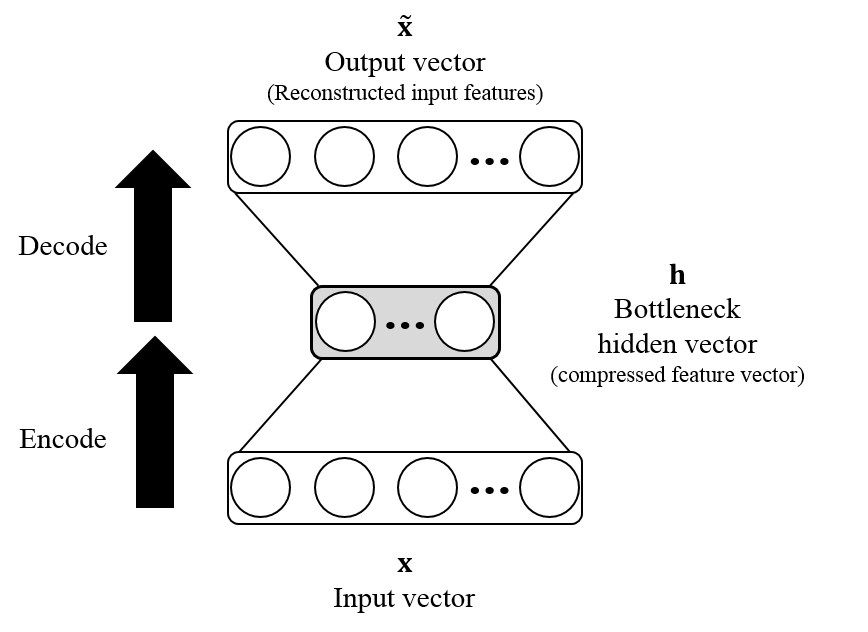}
    \caption{Structure of autoencoder model.}
    \label{fig:AE}
\end{figure*}

In the past several years, bottleneck features have been widely used in many tasks, such as speech recognition \cite{Grzl}-\cite{Sainath}, audio classification \cite{Zhang}, \cite{Seongkyu}, speech synthesis \cite{Wu} and speaker recognition \cite{Yaman}. The bottleneck features are generated from a multi-layer perceptron (MLP) or deep neural network (DNN) with a middle bottleneck layer having small number of hidden units compared to the other hidden layers. This special hidden layer creates a constriction in the network to compress the task-related (classification or regression) information into a low dimensional representation. Therefore, the bottleneck features can be considered as nonlinear transformation and dimensionality reduction of the input features. 

 The bottleneck features can be derived using both unsupervised and supervised method. In unsupervised approach, classically, an autoencoder with one hidden layer trained to predicts input features themselves. The network consists of an encoder and a decoder as shown in Fig. 1. The autoencoder has three layers (input, output and hidden layer). The input vector of autoencoder $\mathbf{x}$ is encoded to hidden vector $\mathbf{h}$ by a nonlinear activation function $\sigma$, using learned weight matrix $\mathbf{W}^{(1)}$ and bias vector $\mathbf{b}^{(1)}$ as follows:
\begin{eqnarray} \label{eq:1}
 \mathbf{h}=\sigma (\mathbf{W}^{(1)} \mathbf{x}+ \mathbf{b}^{(1)}).
\end{eqnarray}
Then, the input vector is decoded from the hidden vector to produce a reconstructed vector $\mathbf{\widetilde{x}}$ using learned weight matrix $\mathbf{W}^{(2)}$ and bias vector $\mathbf{b}^{(2)}$ as follows:
\begin{eqnarray} \label{eq:2}
\mathbf{\widetilde{x}}=\sigma (\mathbf{W}^{(2)} \mathbf{h}+ \mathbf{b}^{(2)}).
\end{eqnarray}
The autoencoder parameter $\theta = (\mathbf{W}^{(1)}, \mathbf{b}^{(1)}), (\mathbf{W}^{(2)}, \mathbf{b}^{(2)})$ is learned using back-propagation algorithm by minimizing the mean square error (MSE) loss as defined:
\begin{eqnarray} \label{eq:3}
L_{\rm{MSE}}(\theta)=\frac{1}{d}\sum_{x \in \rm{D}}{l_{\rm{ MSE}}}(x, \widetilde{x})=\frac{1}{d}\sum_{x \in \rm{D}}\left\| x-\widetilde{x} \right\| ^2.
\end{eqnarray}

\begin{figure*}[tbh]
    \centering
    \includegraphics[height=2.8in]{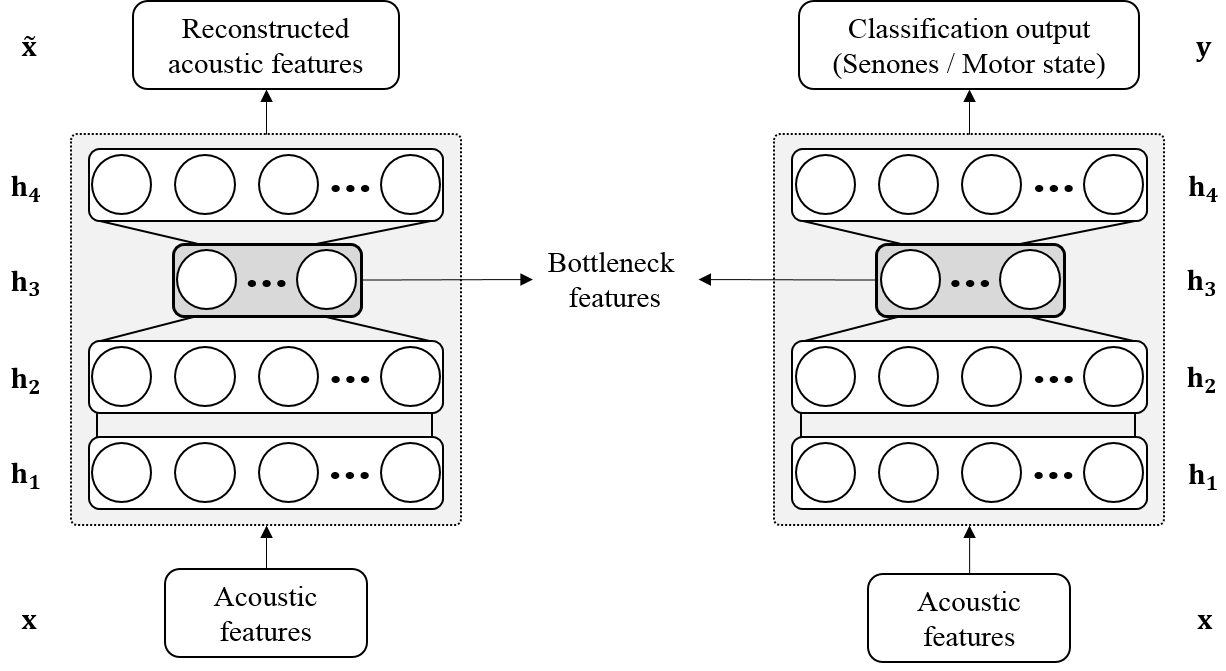}
    \caption{Extracting bottleneck features using unsupervised (left) and supervised (right) methods.}
    \label{fig:SAE_MLP}
\end{figure*}

Further, a stacked autoencoder can be used to extract bottleneck features, which are progressively encoded using successive hidden layers. Firstly, each layer is pre-trained as a shallow autoencoder and the learned hidden layer vector $\mathbf{h}_{l}$ is used to learn the next hidden layer $\mathbf{h}_{l+1}$. Then, fine-tuning on the entire stack of hidden layers is performed using back-propagation algorithm. This allows each hidden layer to provide different levels of representation for the input feature. In stacked autoencoder, the hidden vectors are computed as in (1), for $l=1,2,\ldots,L$:
\begin{eqnarray} \label{eq:4}
\mathbf{h}_{l}=\sigma (\mathbf{W}^{(l)} \mathbf{h}_{l-1}+ \mathbf{b}^{(l)}),
\end{eqnarray}
where $\mathbf{h}_{0}$ is the input vector $\mathbf{x}$ and $L$ denotes the number of hidden layers of stacked autoencoder. 

In the supervised approach, bottleneck features are created by an MLP trained to predict the class label (e.g. phoneme states) as shown in Fig. 2. MLP is feed-forward neural network made of an input layer, output layer, and at least one hidden layer. Usually, for a classification task, the softmax function is adopted to convert the values of arbitrary ranges into a probabilistic representation as defined by
\begin{eqnarray} \label{eq:5}
\sigma(\mathbf{y})=\frac{1}{\sum_{k=0}^{K}{{\rm{exp}}(y_{ k })}}\left[ {\rm{exp}}(y_{ 1 })\cdots {\rm{exp}}(y_{ k }) \right] ^{T},
\end{eqnarray}
where $K$ is the number of elements in $\mathbf{y}$. The learning process attempts to minimize the prediction error $L(x, \widetilde{x})$ with respect to the parameter $\theta = (\mathbf{W}^{(1)}, \mathbf{b}^{(1)}), (\mathbf{W}^{(2)}, \mathbf{b}^{(2)}), \cdots, (\mathbf{W}^{(L)}, \mathbf{b}^{(L)})$. Typically the loss function in MLP is the cross entropy error function \cite{CE}. The supervised method can create a valuable information for classification task. These bottleneck features provide more effective information while preserving enough information of the original input features. 

\section{Proposed methods}

\subsection{Motor on/off state data}

\begin{figure*}[tbh]
    \centering
    \includegraphics[height=2.5in]{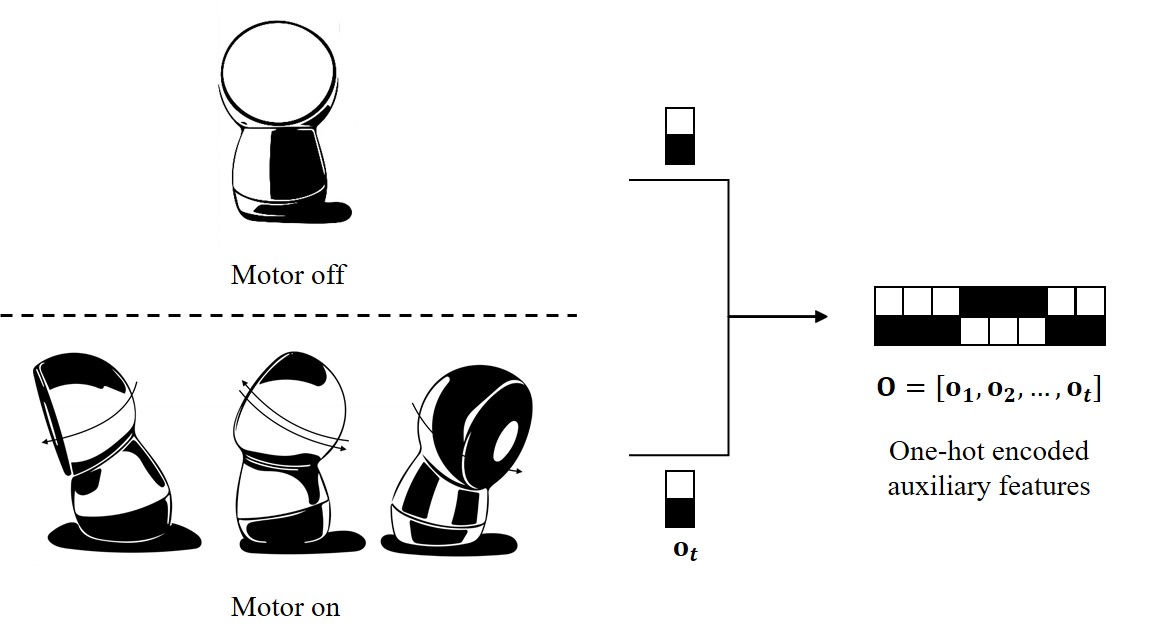}
    \caption{Extracting auxiliary features from the motor on/off state information (``Motor off'' state with fan noise only and ``Motor on'' state with additional movement noise).}
    \label{fig:motor}
\end{figure*}

Since instantaneous motor state information of the robot is intuitively useful for handling the ego-noise problem, in this paper, we fuse acoustic information obtained from spoken utterances and the motor state information brought by the instantaneous motor on/off state data into a single framework. To do this, we propose a method to use the motor state data as auxiliary features, trains the bottleneck features. The motor data derived from the robot can be classified into a basic operation state in which only the fan and the motor are turned on (``motor off''), and a motion state in which the robot shakes its head or body according to the human command (``motor on''). Our robot transmits auxiliary information with the basic operation state as ``state off'' and the other state as ``state on''. This auxiliary feature can be observed at each frame and contains instantaneous internal state information of the humanoid robot. The concatenated input with conventional spectral features $\mathbf{x'}=[\mathbf{x}; \mathbf{o}]$ is used for bottleneck feature learning.


\subsection{Extracting motor state dependent bottleneck features}

\begin{figure*}[tbh]
    \centering
    \includegraphics[height=3.8in]{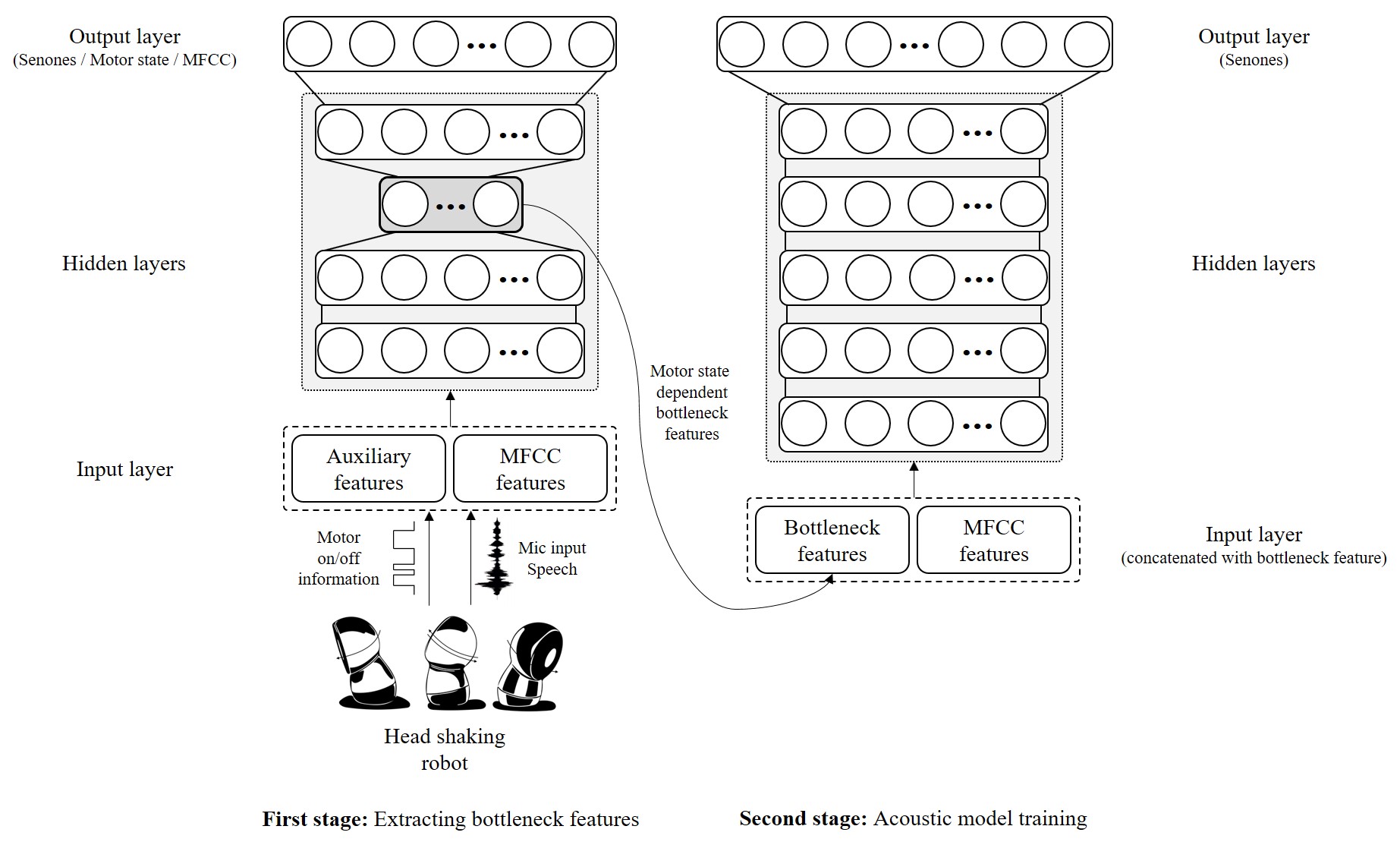}
    \caption{Framework of the proposed ASR system employing bottleneck features}
    \label{fig:Structure}
\end{figure*}

The elementary question is to how we fuse the motor data into the conventional acoustic features. Herein, we propose to extend the bottleneck feature-based ASR method, motivated in \cite{ego-noise5, DongYu}. The key novelty is to learn the motor state dependent bottleneck features based on additional instantaneous motor data. In this work, 4 hidden layers of structure including bottleneck layer was selected on both unsupervised and supervised methods as shown in left-hand of Fig. 4. The MFCC and auxiliary features are concatenated at each frames and consecutive frames are used to train the bottleneck features. From this, the motor on/off state data can be encoded to more effective representation.

\subsection{Acoustic model training}

Fig. 4 illustrates the overall architecture of the proposed ASR system that employs motor state dependent bottleneck features. The left-hand network is the bottleneck network to extract ego-noise adaptive bottleneck features first. Then, the bottleneck features are stacked alongside the spectral features as input to the right-hand network in order to train the acoustic model. The concatenated features contain the motor state information that is needed to build ego-noise robust speech recognition system. We will investigate the performance of various such system configurations in the next section.


\section{Experiments and Results}

\subsection{Corpus Description}

\captionsetup[table]{labelformat={default},labelsep=period,name={Table}}
\begin{table}[]
\centering
\caption{Hardware specifications of JIBO \cite{JIBO}.}
\label{my-label}
\begin{tabular}{|l|l|}
\hline
\textbf{Hardware} & \textbf{Specifications}        \\ \hline
Sensors           & 360 degrees sound localization \\ \hline
Movement          & 3 full-revolute axes           \\ \hline
Sound             & 2 premium speakers             \\ \hline
Processor         & High-end ARM-based mobile      \\ \hline
\end{tabular}
\end{table}

In order to evaluate the proposed approach, we conducted experiments with a JIBO humanoid robot which has 3 full-revolute axes \cite{JIBO}. A brief specification of the robot is introduced in Table I. We consider a scenario in which humans interact with a robot while the robot shakes his head. To simulate noisy environments in the humanoid robot, we recorded ego-noises using the single microphone located at the front side of the head. These noise signals involve two types: fan noise and movement noise. The mixing is conducted at various signal-to-noise ratio (SNR) levels including 5 dB, 10 dB, 15 dB and 20 dB, depending on the distance between the speaker and robot. These mixtures were then used to train and evaluate the ego-noise robust ASR algorithms described above. Our experiments were conducted on the TIMIT database \cite{TIMIT} divided into three subsets: 3969 utterances as training set, 400 utterances as development set, and 192 utterances as testing set. The waveform sampling rate of the corpus and the recorded noises was 16 kHz. We then measured the proposed algorithm in terms of phoneme error rate (PER) under the aformention environments. 

\subsection{Experimental Setup}

In our experiments, the Kaldi toolkit \cite{Kaldi} was utilized to train the bottleneck network and the acoustic model. The systems implemented and used for comparison in our experiments are as follows:


\emph{1) DNN }(\emph{MFCC}): A baseline system using no motor data but conventional spectral features, obtained from the spoken utterances, as the input features for training acoustic model. 

\emph{2) DNN }(\emph{MFCC + motor data}): A second baseline system using auxiliary features in addition to the conventional spectral features as the input features. The auxiliary features were given by the one-hot representation of the motor on/off state information.

\emph{3) BN-DNN-PHN }(\emph{MFCC + BN-PHN}): The proposed system using motor state dependent bottleneck features as auxiliary features. The ego-noise adaptive features, rather than the simple one-hot representation of motor data, were combined with conventional spectral features and used to train acoustic model. In order to extract ego-noise adaptive features, the one-hot encoded motor state data and the spectral features were utilized for the input features and phoneme (PHN) states were employed for the output features.

\emph{4) BN-DNN-MS }(\emph{MFCC + BN-MS}): Same as BN-DNN-PHN except the output features of the bottleneck network were one-hot representation of motor state (MS) data. It was trained to classify the motor on/off state of the robot.

\emph{5) BN-DNN-MFCC }(\emph{MFCC + BN-MFCC}): Same as BN-DNN-PHN except the output features of the bottleneck network were original MFCC features. It was considered as an autoencoder.





All the systems described above employed the same structure of acoustic model having 5 hidden layers each of which has 512 hidden units. The rectified linear unit (ReLU) activation functions were used in the lower layers, and a softmax function at the output layer. For the conventional spectral features, 13 dimensional MFCC features were extracted using 25ms analysis window with 10 ms frame shift. As for the input of acoustic model, in the baseline systems, the MFCC features stacked with 11 adjacent frames were used and the additional one-hot representation of the motor data was used for the second baseline system. ($13 \times 11 = 143$-dim. for the first baseline system and $15 \times 11 = 165$-dim. for the second respectively.)

In the proposed systems, the additional bottleneck networks with 4 hidden layers were trained separately. To compare the effect of the various bottleneck features, experiments were performed with different output features, bottleneck layer dimensions and bottleneck layer positions. Firstly, the PHN, MS label and original MFCC features were compared as output features. The sigmoid and tanh activation functions were used for classification and regression task, respectively. Also, the bottleneck sizes of 40-dim and 80-dim were compared. Finally, we varied the placement of the bottleneck layer from the bottom hidden layer (position 1) to the top hidden layer (position 4). For all the bottleneck networks, stacked MFCC and auxiliary features ($(13 + 2) \times 11 = 165$-dim.) were used for input. Then, the extracted motor state dependent bottleneck features were combined with the spectral features again and used to train the acoustic model as shown in Fig. 4.

\subsection{Experimental Results and Analysis}

\captionsetup[table]{labelformat={default},labelsep=period,name={Table}}
\begin{table}[]
\centering
\caption{Performance (PER in \%) comparison when only fan noise exists ("motor off state"). The PHN, MS and MFCC indicate output of the bottleneck network. (PHN: phoneme, MS: motor state, and MFCC: mel-frequency cepstral coefficients). The BN and BN2 indicate 40 and 80 dimensional bottleneck features, respectively.}
\label{my-label}
\begin{tabular}{lccccc}
\hline
               & \multicolumn{5}{c}{PER (\%)}                                            \\ \cline{2-6} 
               & 5 dB & 10 dB & 15 dB & 20 dB & Avg. \\ \hline
\multicolumn{6}{c}{Baseline}                                                                      \\ \hline
MFCC           & 31.1          & 29.1           & 27.9           & 26.9           & 28.8          \\
MFCC + MS      & 31.2          & 28.7           & 26.5           & 25.7           & 28.0         \\ \hline
\multicolumn{6}{c}{Proposed}                                                                      \\ \hline
MFCC + BN-PHN   & 28            & 25.6           & 24.2           & 24.2           & \textbf{25.5} \\
MFCC + BN-MS   & 30.9          & 27.6           & 25.9           & 25.5           & 27.5          \\
MFCC + BN-MFCC   & 30            & 27.6           & 26.7           & 26.1           & 27.6          \\
MFCC + BN2-PHN & 28.4          & 25.6           & 24.6           & 24.1           & \textbf{25.7}   \\
MFCC + BN2-MS  & 29.7          & 27.2           & 25.9           & 25.8           & 27.1            \\
MFCC + BN2-MFCC  & 30.5          & 27.9           & 26.5           & 25.6           & 27.6          \\ \hline
\end{tabular}
\end{table}


\captionsetup[table]{labelformat={default},labelsep=period,name={Table}}   
\begin{table}[]
\centering
\caption{Performance (PER in \%) comparison when there is an additional movement noise in addition to the fan noise (``motor on state").}
\label{my-label}
\begin{tabular}{lccccc}
\hline
               & \multicolumn{5}{c}{PER (\%)}                                            \\ \cline{2-6} 
               & 5 dB & 10 dB & 15 dB & 20 dB & Avg. \\ \hline
\multicolumn{6}{c}{Baseline}                                                                      \\ \hline
MFCC           & 31.6          & 29.1           & 27.7           & 26.8           & 28.8          \\
MFCC + MS      & 31.4          & 28.5           & 26.8           & 26.1           & 28.2          \\ \hline
\multicolumn{6}{c}{Proposed}                                                                      \\ \hline
MFCC + BN-PHN   & 28.7          & 26.1           & 24.7           & 23.5           & \textbf{25.8} \\
MFCC + BN-MS   & 31            & 27.6           & 26             & 25.5           & 27.5          \\
MFCC + BN-MFCC   & 30.8          & 27.9           & 26.7           & 26.2           & 27.9          \\
MFCC + BN2-PHN & 29.2          & 26             & 24.7           & 23.8           & \textbf{25.9} \\
MFCC + BN2-MS  & 29.7          & 27.2           & 26.1           & 25.6           & 27.2          \\
MFCC + BN2-MFCC  & 30.3          & 27.8           & 26.7           & 25.4           & 27.6         \\ \hline
\end{tabular}
\end{table}


Table I and II present the PER results on the ``motor on" and ``motor off" state, respectively. It is worth noting first that the motor state data shows a better recognition performance on the both states. In particular, the auxiliary features, generated by using the bottleneck network, yielded superior performance when compared to one-hot encoded vectors. It indicates that the bottleneck network can create more valuable representation of the motor state data by fusing along with the spectral features and being compressed.

For comparison of the proposed algorithms, the PER is reported for each output features. The results show that the phoneme states are appropriate as target features. The model with bottleneck features predicting phonemes (BN-PHN) achieved relative PER reduction of 11.5\% and 10.4\% over the baseline model using no motor data on the ``motor off" and ``motor on" states respectively. Furthermore, we examined the effect of the bottleneck feature size and it did not show any significant performance differences. Therefore, considering the computational complexity, 40 dimensional bottleneck feature is suitable for extracting motor state dependent bottleneck features.

In addition, the effect of bottleneck layer position is presented in Fig. 5 on both (a) 40 and (b) 80 dimensional bottleneck hidden layer experiments. From the results, It is evident that the middle (second or third) layer is reasonable for the phoneme class and the first hidden layer is moderate for the others.

\begin{figure*}[tbh]
    \centering
    \begin{subfigure}[b]{0.5\textwidth}
        \centering
        \includegraphics[height=2.4in]{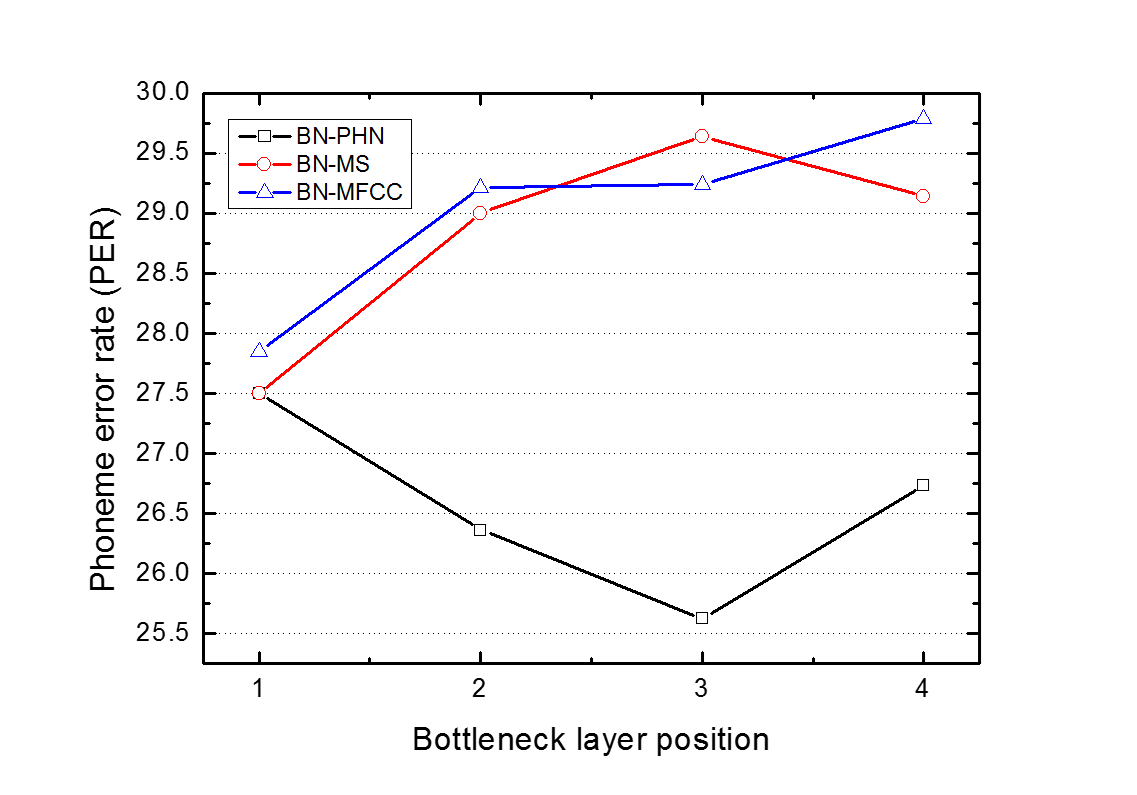}
        \caption{}
    \end{subfigure}%
    ~ 
    \begin{subfigure}[b]{0.5\textwidth}
        \centering
        \includegraphics[height=2.4in]{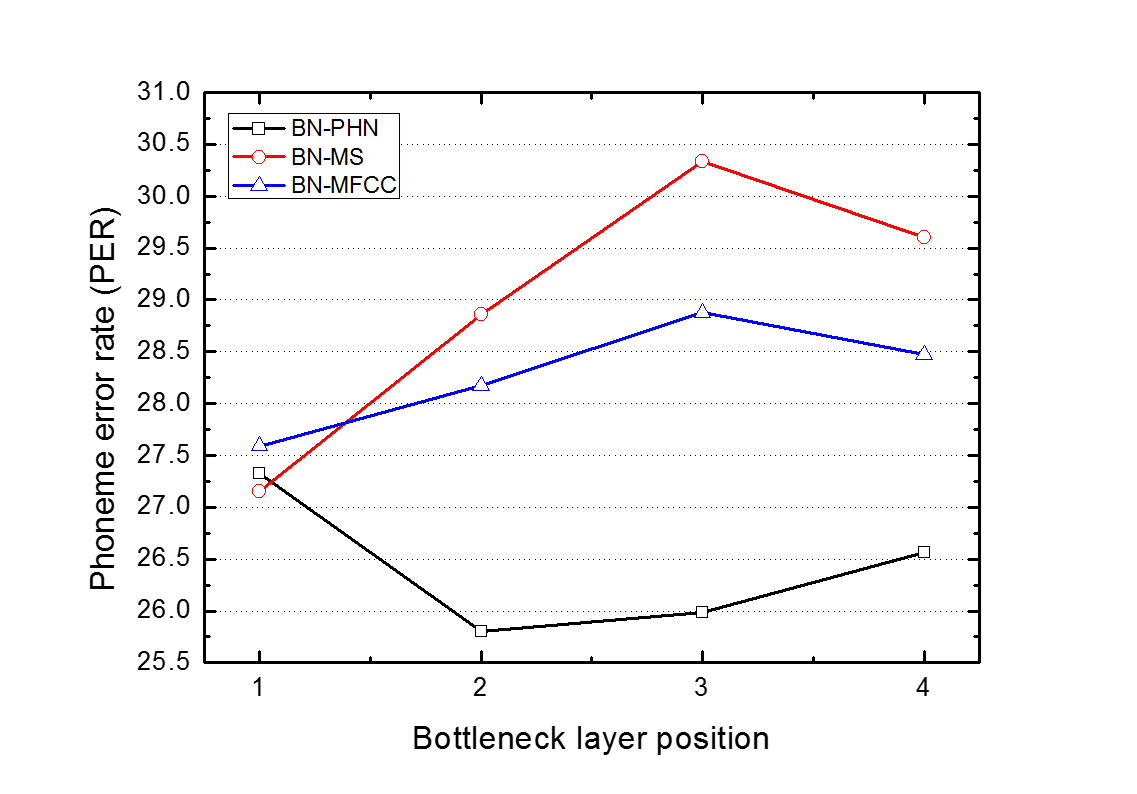}
        \caption{}
    \end{subfigure}
    \caption{PER as a function of the position of the bottleneck layer (The 40-dimensional bottleneck layer in (a) and the 80-dimensional bottleneck layer in (b)).}
\end{figure*}

\section{Conclusion}
In this paper, we proposed a novel method to incorporate the instantaneous motor on/off state information into a ego-noise robust ASR system, which results in a better performance than exploiting no motor data. For this, we employed a bottleneck network to create motor state dependent bottleneck features to effectively integrate the motor data along with conventional speech signals. These ego-noise adaptive bottleneck features provide a significant improvement than one-hot encoded motor state features. We investigated the effect of output features of the bottleneck network and shown that the phoneme states classification output is most effective to extract ego-noise adaptive bottleneck features. Additionally, we compared the effect of the bottleneck layer position and concluded that the middle (second or third) layer is reasonable for the phoneme class and the first hidden layer is moderate for the others. From the experimental results, we concluded that the robot's instantaneous motor state information is advantageous for human-robot communication. In particular, the bottleneck network can generate more valuable representation of motor data than one-hot encoding method. In a future work, more varied states of the robot will be considered as motor data, e.g. walking state, right/left arm rotating state, head shaking state, or multiple state.


\ifCLASSOPTIONcaptionsoff
  \newpage
\fi


\end{document}